%% file: CoSi_261113-1.tex
\documentclass[%
reprint,
superscriptaddress,
amsmath,amssymb,
prb,
longbibliography,
]{revtex4-2}

\input{revpackage.tex}

\usepackage{multirow}

\begin{document}

\title{Raman optical activity and helicity-dependent phonon splitting in chiral CoSi}

\author{Yuki Suganuma}
\affiliation{Department of Physics, Institute of Science Tokyo, Tokyo 152-8551, Japan}

\author{Gakuto Kusuno}
\affiliation{Department of Physics, Institute of Science Tokyo, Tokyo 152-8551, Japan}

\author{Kohei Miyazaki}
\affiliation{Department of Physics, Institute of Science Tokyo, Tokyo 152-8551, Japan}

\author{Hikaru Watanabe}
\affiliation{Graduate School of Engineering, Hokkaido University, Sapporo 060-8628, Japan}
\affiliation{Department of Physics, The University of Tokyo, Tokyo 113-0033, Japan}

\author{Rikuto Oiwa}
\affiliation{Graduate School of Science, Hokkaido University, Sapporo 060-0810, Japan}

\author{Ryotaro Arita}
\affiliation{Department of Physics, The University of Tokyo, Tokyo 113-0033, Japan}
\affiliation{Center for Emergent Matter Science, RIKEN, Wako 351-0198, Japan}

\author{Satoshi Iwasaki}
\affiliation{Research Institute for Interdisciplinary Science, Okayama University, Okayama 700-8530, Japan}

\author{Yoshiki Yasuoka}
\affiliation{Department of Physics and Electronics, Graduate School of Engineering, Osaka Metropolitan University, Osaka 599-8531, Japan}

\author{Yusuke Kousaka}
\affiliation{Department of Physics and Electronics, Graduate School of Engineering, Osaka Metropolitan University, Osaka 599-8531, Japan}

\author{Yoshihiko Togawa}
\affiliation{Department of Physics and Electronics, Graduate School of Engineering, Osaka Metropolitan University, Osaka 599-8531, Japan}
\affiliation{Quantum Research Center for Chirality, Institute for Molecular Science, Okazaki 444-8585, Japan}

\author{Takuya Satoh}
\email{satoh@phys.sci.isct.ac.jp}
\affiliation{Department of Physics, Institute of Science Tokyo, Tokyo 152-8551, Japan}
\affiliation{Quantum Research Center for Chirality, Institute for Molecular Science, Okazaki 444-8585, Japan}

\date{\today}

\begin{abstract}
Circularly polarized Raman scattering in solids provides access to two distinct phenomena: Raman optical activity (ROA), an intensity difference between circular polarization channels, and helicity-dependent frequency splitting of chiral phonons. We observe both in the chiral topological semimetal CoSi and show that they arise through two different symmetry channels, each rendering phonon components with opposite crystal angular momenta inequivalent. Doubly degenerate $E$ phonons exhibit a large ROA, with an approximately fourfold intensity contrast between the two cross-circular channels. The sign of this ROA reverses between the front and back $\{111\}$ surfaces, while its strong suppression on $\{110\}$ supports the assignment to the axial electric toroidal octupole. By contrast, triply degenerate $T$ phonons show helicity-dependent frequency splittings, whose mode-dependent signs are reproduced by the calculated dispersion of branches with opposite crystal angular momenta, thus identifying structural chirality as their source. First-principles calculations yield a pronounced $E$-mode intensity asymmetry in the resonant regime, reproduce the mode-dependent signs of the $T$-mode splittings, and account for their magnitudes within approximately 40\%. These results establish circularly polarized Raman scattering as a symmetry- and angular-momentum-resolved probe of chiral and axial multipolar symmetry in solids.
\end{abstract}

\maketitle

\section{Introduction}

Raman optical activity (ROA), the difference in Raman scattering intensity between circular polarization channels, was established in molecular systems, where natural (nonmagnetic) ROA reflects molecular chirality \cite{BarronBook, Barron1973}. ROA has since been reported in solids, including the charge-density-wave material 1T-TaS$_2$ \cite{Lacinska2022, Yang2022} and the chiral and polar compound NiCo$_2$TeO$_6$ \cite{Martinez2025}. More recently, pronounced ROA has been shown to arise from axial order in the ferroaxial systems NiTiO$_3$ \cite{Kusuno2026NiTiO3} and 4Hb-TaS$_2$ \cite{Barantani2026}, and from higher-rank multipolar symmetry in pyrite FeS$_2$ \cite{Suganuma2026Pyrite}, alongside symmetry-based analyses that classify circular Raman responses by the broken symmetries and the associated selection rules \cite{Watanabe2025PRB, Watanabe2025arXiv}.

In parallel, circularly polarized Raman spectroscopy has revealed a different phenomenon in chiral crystals: the frequency splitting of degenerate phonon modes \cite{Zhang2025Review}. Phonons carrying crystal angular momentum (CAM, also called pseudoangular momentum \cite{Streib2021}) were first observed in two-dimensional lattices \cite{Zhu2018}. Following Ref.~\cite{Juraschek2025}, we call phonons carrying angular momentum \cite{Zhang2014} or CAM \cite{Bozovic1984, Zhang2015, Zhang2022} \emph{axial}, and reserve \emph{chiral} for phonons with a finite $\mathbf{L}\cdot\mathbf{q}$ \cite{Ishito2023Te}, the pseudoscalar formed from the phonon angular momentum $\mathbf{L}$ and the wave vector $\mathbf{q}$, which vanishes unless all improper rotational symmetries are absent. Such chiral phonons have been observed in three-dimensional chiral crystals including $\alpha$-HgS \cite{Ishito2023HgS}, Te \cite{Ishito2023Te}, $\alpha$-SiO$_2$ \cite{Oishi2024, Ueda2023}, and CrSi$_2$ \cite{Kusuno2026CrSi2}.

These two families of circular Raman responses have been primarily discussed separately because they originate from different symmetry mechanisms and appear in different observables, namely intensity asymmetry and frequency splitting. Here, we observe both in CoSi, a chiral topological semimetal \cite{Rao2019, Xu2020} that simultaneously possesses structural chirality and an axial octupolar degree of freedom \cite{Hayami2024, Hayami2018}. Its electronic system also exhibits pronounced helicity-dependent nonlinear optical responses \cite{Ni2021, Chan2024}. We separate the two Raman responses into the two symmetry channels through which states with opposite CAM are rendered inequivalent.

We find that doubly degenerate $E$ phonons exhibit a large ROA, which reverses sign between opposite $\{111\}$ surfaces and is suppressed on $\{110\}$, whereas triply degenerate $T$ phonons exhibit helicity-dependent frequency splittings, whose mode-dependent signs are reproduced by first-principles calculations. 

\section{Methods}
\label{sec:methods}

\subsection{Crystal structure and growth}

\begin{figure}
\centering
\includegraphics[width=0.9\hsize]{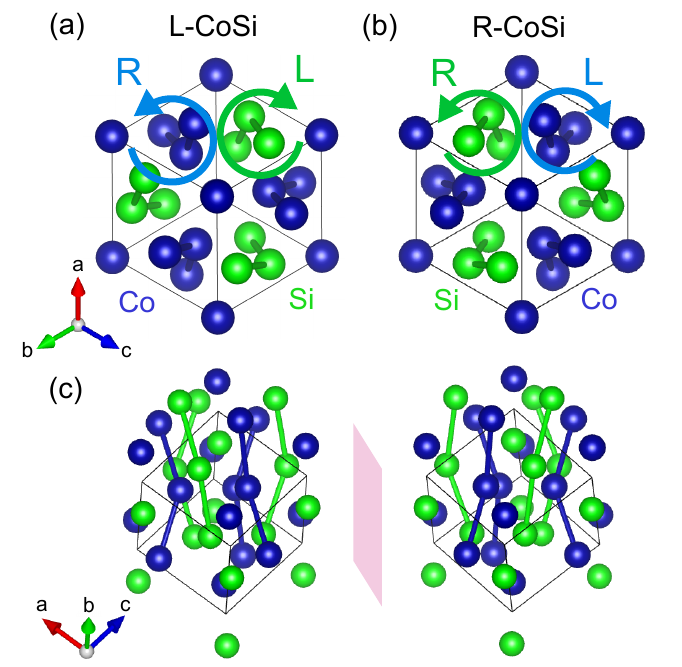}
\caption{Crystal structures of CoSi. (a) Left-handed and (b) right-handed enantiomers viewed along $[111]$, visualized using VESTA \cite{Momma2011vesta}. Curved arrows give the sense of the Co and Si screws (R, right-handed; L, left-handed). The crystal handedness follows that of the Si helices \cite{Kousaka2022}, so L-CoSi has a right-handed Co helix. These labels are distinct from the circular-polarization labels L and R. (c) The same two enantiomers viewed along a different crystallographic direction. The shaded plane indicates the mirror relating them. Blue and green spheres represent Co and Si atoms, respectively.}
\label{CoSiatom}
\end{figure}

CoSi crystallizes in the chiral cubic B20 structure with space group $P2_13$ (No.~198), whose point group is the noncentrosymmetric cubic group $23$ \cite{Kousaka2022}. Devoid of mirror and inversion symmetries, it occurs as left- and right-handed enantiomers [Figs.~\subref{CoSiatom}{a}--\subref{CoSiatom}{c}]. The zone-center optical phonons are classified into irreducible representations $\Gamma_{\mathrm{optical}} = 2A + 2E + 5T$, where $A$ is nondegenerate, $E$ doubly degenerate, and $T$ triply degenerate; all are Raman active.

Single crystals of CoSi were grown from composition-gradient feed rods by a laser-diode-heated floating-zone method in which the crystallographic chirality is inherited from an enantiopure seed crystal~\cite{Kousaka2022}. The handedness was assigned by single-crystal x-ray absolute structure analysis (XtaLABmini II, Rigaku Corporation). The two $\{111\}$ faces are referred to here as the front and back surfaces. Crystals were oriented and cut to expose $\{111\}$ and $\{110\}$ surfaces. All measurements reported here were performed on a left-handed crystal, L-CoSi, which exhibited the highest crystalline quality.

\subsection{Circularly polarized Raman spectroscopy}
\label{sec:experiment}

Circularly polarized Raman spectra were measured at room temperature ($295$~K) in backscattering at normal incidence, using primarily a $785$-nm ($1.58$-eV) excitation laser. A $633$-nm ($1.96$-eV) laser was used for the excitation-wavelength dependence. Left (L) and right (R) polarizations are defined by the sense of rotation of the electric field in the sample plane, independent of the propagation direction [see the inset of Fig.~\subref{CoSiraman}{a}]. All four configurations LR, RL, LL, and RR were recorded, the first and second labels denoting the incident and scattered polarizations. Peak positions and integrated intensities were obtained from Voigt fits, and systematic shifts between the LR and RL configurations were minimized by recording the two configurations under otherwise identical conditions. Details of the optical setup are given in the Supplemental Material \cite{SM} (Sec.~S1).

\subsection{First-principles calculations}
\label{sec:calc}

Phonon dispersions and Raman spectra were computed based on density functional perturbation theory~\cite{Baroni2001} as implemented in \textsc{quantum espresso} \cite{Giannozzi2009-wa, Giannozzi2017-do}, using scalar-relativistic norm-conserving pseudopotentials generated with the optimized norm-conserving Vanderbilt scheme~\cite{PhysRevB.88.085117} and the Perdew--Burke--Ernzerhof exchange-correlation functional \cite{Perdew1996-oq}. Phonon properties were analyzed using \textsc{phonopy} \cite{phonopy-phono3py-JPCM, phonopy-phono3py-JPSJ}, and the Raman intensities were evaluated from Wannier-interpolated electronic states using \textsc{wannier90} \cite{Pizzi2020-uz}, \textsc{epw} \cite{Lee2023-yc}, and \textsc{qr2code} \cite{HUANG2026110005}. For phonon mode $\nu$ and incident photon energy $E_{\mathrm{L}}$, the calculated polarization-resolved Raman amplitude is
\begin{equation*}
M_{\nu}(E_{\mathrm{L}}) = \mathbf{e}_s^{\dagger}\,\hat{\mathcal{R}}_{\nu}(E_{\mathrm{L}})\,\mathbf{e}_i, \qquad I_{\nu}(E_{\mathrm{L}}) \propto |M_{\nu}(E_{\mathrm{L}})|^{2},
\end{equation*}
where $\hat{\mathcal{R}}_{\nu}(E_{\mathrm{L}})$ is the complex resonant Raman tensor for the one-phonon process that is constructed from electron--photon and electron--phonon matrix elements. For the two $E$ components we write $\nu = \pm$. Further details are provided in the Supplemental Material \cite{SM} (Sec.~S2). These calculations were performed within the electric-dipole approximation.

\subsection{Circular-polarization selection rules}
\label{sec:selection}

We write the optical fields with the time dependence $e^{-\mathrm{i}\omega t}$, so that a unit polarization vector $\mathbf{e}$ carries the photon spin $\mathrm{Im}(\mathbf{e}^{*}\times\mathbf{e})\cdot\hat{\mathbf{n}}$ in units of $\hbar$ along the incident propagation direction $\hat{\mathbf{n}} \parallel [111]$, which serves as the reference for both beams. Right- and left-circular polarizations are named by the sign of that spin, positive and negative respectively, as in the convention in which a right-circularly polarized photon carries positive angular momentum along its propagation direction \cite{Yariv2006}. The angular momentum transferred to the crystal along $[111]$ is then $\Delta\sigma = \sigma_{\mathrm{i}} - \sigma_{\mathrm{s}}$ with $\sigma = -1$ for L and $\sigma = +1$ for R, where $i$ and $s$ label the incident and scattered photons. Because of the threefold rotational symmetry about $[111]$, each phonon mode is labeled by its CAM $m = 0, \pm 1$, defined modulo 3 by $C_3^{+}\mathbf{u} = e^{-2\pi\mathrm{i}m/3}\,\mathbf{u}$ for the phonon eigenvector $\mathbf{u}$, the sign convention of Refs.~\cite{Zhang2015,Ishito2023HgS}, with $C_3^{+}$ the rotation $x \to y \to z \to x$ about $[111]$. Throughout, $m$ is the CAM of the phonon created in the Stokes process, and the normal coordinate that creates it has the conjugate $C_3^{+}$ eigenvalue (Appendix). The cross-circular configurations select $\Delta\sigma = \mp 2 \equiv \pm 1 \pmod 3$, so that, for the Stokes process, LR probes phonons with $m = +1$ and RL probes $m = -1$, whereas the parallel-circular configurations LL and RR probe $m = 0$ \cite{Ishito2023HgS}. Table~\ref{tab:selection} summarizes these selection rules and the resulting intensities. The derivation from the Raman tensors of class $23$ is given in Appendix.

The CAM $m$ is defined with respect to the $[111]$ axis of the crystal, whereas the polarization labels follow the fixed laboratory convention described in Sec.~\ref{sec:experiment}. Probing the opposite $\{111\}$ surface reverses the propagation direction relative to the crystal axes and hence the correspondence between the laboratory labels and crystal-frame helicities.

\begin{table}[t]
\caption{Circular-polarization selection rules for light propagating along $[111]$. Here $m$ is the CAM selected in the Stokes process. The Raman tensor coefficients $a$ and $c$ and the $E$-mode susceptibilities $\chi_\pm$ are defined in Appendix.}
\begin{ruledtabular}
\begin{tabular}{cccccc}
Config. & $\sigma_{\mathrm{i}}$ & $\sigma_{\mathrm{s}}$ & $\Delta\sigma$ & $m$ & Intensity \\ \hline
LR & $-1$ & $+1$ & $-2$ & $+1$ & $A:0$, $E:|\chi_+|^2$, $T:4c^2/3$ \\
RL & $+1$ & $-1$ & $+2$ & $-1$ & $A:0$, $E:|\chi_-|^2$, $T:4c^2/3$ \\
LL & $-1$ & $-1$ & $0$ & $0$ & $A:a^2$, $E:0$, $T:c^2/3$ \\
RR & $+1$ & $+1$ & $0$ & $0$ & $A:a^2$, $E:0$, $T:c^2/3$ \\
\end{tabular}
\end{ruledtabular}
\label{tab:selection}
\end{table}

\section{Results}
\label{sec:results}

\subsection{Raman spectra and mode assignments}

\begin{figure*}
\includegraphics[width=\hsize]{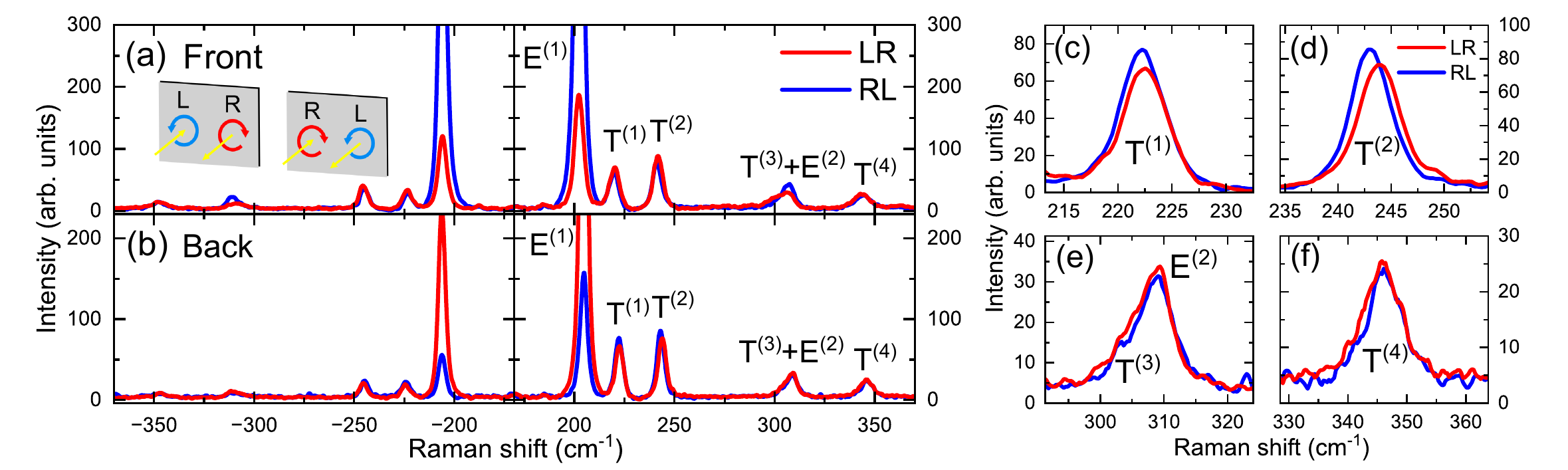}\vspace{-10pt}
\caption{(a),(b) Anti-Stokes and Stokes Raman spectra of L-CoSi measured on  (a) front and (b) back $\{111\}$ surfaces at an excitation wavelength of 785 nm in cross-circular polarization configurations. Inset of (a): schematic illustration showing cross-circular configurations LR and RL. (c)--(f) Enlarged spectra of the back $\{111\}$ surface showing circular-polarization dependence of (c) $T^{(1)}$, (d) $T^{(2)}$, (e) $T^{(3)}+E^{(2)}$, and (f) $T^{(4)}$ modes.}
\label{CoSiraman}
\end{figure*}

Figures~\subref{CoSiraman}{a} and \subref{CoSiraman}{b} show the circularly polarized Raman spectra of L-CoSi measured in the cross-circular configurations on the front and back $\{111\}$ surfaces of the same crystal. Several phonon modes are observed in the measured range, corresponding to optical phonons near the $\Gamma$ point. Spectra recorded in the parallel-circular configurations, in which the $A$ modes appear as required by Table~\ref{tab:selection}, are given in the Supplemental Material \cite{SM} (Sec.~S3).

Table~\ref{tab:modes} shows a comparison of the observed frequencies with previous linearly polarized Raman data \cite{Racu2007} and the present calculations. The agreement supports the mode assignments used in this study. A small systematic overestimation of the calculated frequencies is common in first-principles phonon calculations and does not affect the ordering or symmetry assignments.

\begin{table}[t]
\caption{Frequencies of Raman-active phonons of CoSi at the $\Gamma$ point, in cm$^{-1}$. Previous Raman data are taken from the $x = 1$ composition of Ref.~\cite{Racu2007}.}
\begin{ruledtabular}
\begin{tabular}{cccc}
 & \multicolumn{2}{c}{Experiment} & \\
\cline{2-3}
Mode & This work & Ref.~\cite{Racu2007} & Calculation \\ \hline
$A^{(1)}$ & 186 & 184 & 187 \\
$E^{(1)}$ & 205 & 203 & 209 \\
$T^{(1)}$ & 222 & 220 & 227 \\
$T^{(2)}$ & 244 & 242 & 247 \\
$T^{(3)}$ & 304 & 302 & 310 \\
$E^{(2)}$ & 309 & 309 & 313 \\
$T^{(4)}$ & 346 & 344 & 350 \\
$A^{(2)}$ & 364 & 362 & 363 \\
$T^{(5)}$ & --- & 419 & 422 \\
\end{tabular}
\end{ruledtabular}
\label{tab:modes}
\end{table}

\subsection{Optical activity of \texorpdfstring{$E$}{E} modes}
\label{sec:roa}

A pronounced difference in Raman intensity between the LR and RL configurations is observed for the doubly degenerate $E$ modes [Figs.~\subref{CoSiraman}{a} and \subref{CoSiraman}{b}]. The $E^{(1)}$ mode at 205~cm$^{-1}$ shows the largest effect, with one circular polarization channel substantially more intense than the other. We quantify this asymmetry by the ROA factor \cite{Kusuno2026NiTiO3, Suganuma2026Pyrite}
\begin{equation}
g_{\mathrm{ROA}} = \frac{2(I_{\mathrm{LR}} - I_{\mathrm{RL}})}{I_{\mathrm{LR}} + I_{\mathrm{RL}}} ,
\label{eq:groa}
\end{equation}
where $I_{\mathrm{LR}}$ and $I_{\mathrm{RL}}$ are the integrated intensities obtained from the Voigt fits, so that $|g_{\mathrm{ROA}}| \leq 2$.
For the $E^{(1)}$ mode, we obtained $g_{\mathrm{ROA}} = -1.27$ (front) and $+1.18$ (back) for the Stokes process, and $-1.31$ (front) and $+1.25$ (back) for the anti-Stokes process.
These values correspond to ratios of the stronger to the weaker cross-circular channel ranging from $3.9$ to $4.8$.

Two features characterize this asymmetry. First, the sign of $g_{\mathrm{ROA}}$ is the same for the Stokes and anti-Stokes processes. This contrasts with magnetic ROA, in which the sign reverses between the two processes \cite{Barron1982, Cenker2021, Watanabe2025PRB}. Second, the sign reverses between the front and back $\{111\}$ surfaces of the same crystal. Under reversal of the measurement direction, the two cross-circular channels are exchanged, so that $g_{\mathrm{back}}(q) = -g_{\mathrm{front}}(-q)$ (Sec.~\ref{sec:discussion}). Spatial maps over the measured area confirm that this reversal is uniform and not a local artifact [Supplemental Material \cite{SM} (Sec.~S4)].

\begin{figure}
\centering
\includegraphics[width=0.85\hsize]{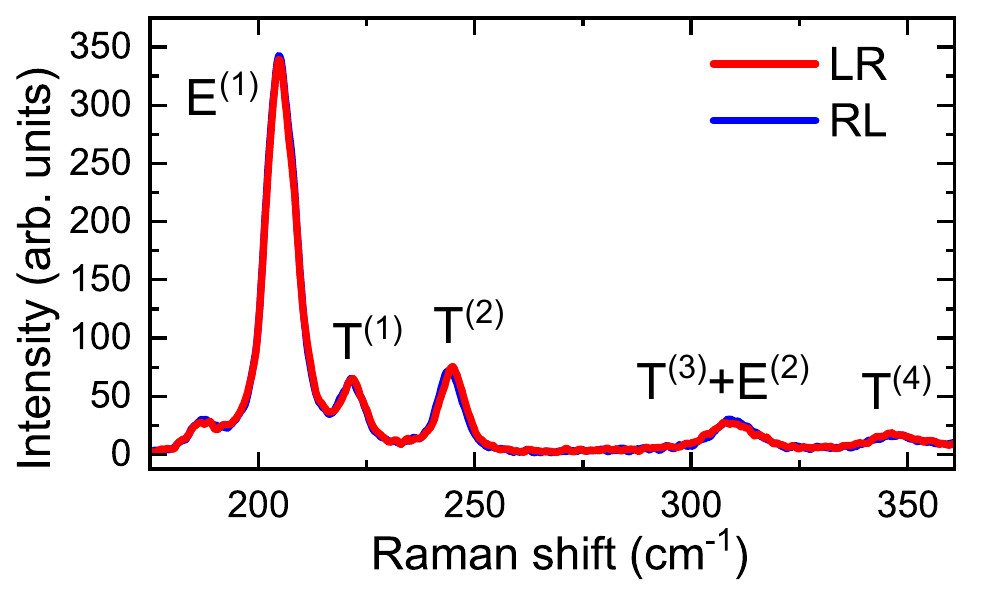}
\caption{Stokes Raman spectra of L-CoSi measured on the $\{110\}$ surface at an excitation wavelength of 785 nm in cross-circular polarization configurations. Compared with the case of $\{111\}$ surfaces, the difference between LR and RL spectra is strongly suppressed.}
\label{fig:110}
\end{figure}

The surface-orientation dependence provides a further test. On the $\{110\}$ surface, the difference between the LR and RL spectra is strongly suppressed (Fig.~\ref{fig:110}). The selection rules described in Sec.~\ref{sec:selection} do not apply here, since they are derived for propagation along $[111]$ (Appendix). The suppression follows instead from the point-group symmetry of the propagation direction itself: a twofold axis of class $23$ is perpendicular to $\langle 110 \rangle$, which forces the parity-even octupolar contribution to the cross-circular response to vanish, as discussed in Sec.~\ref{sec:discussion}.

Finally, the effect depends strongly on the excitation wavelength: the intensity asymmetry is largely suppressed for 633-nm excitation [Supplemental Material \cite{SM} (Sec.~S4)], indicating that the electronic transitions involved in the scattering process play an essential role. This is examined together with the calculations in Sec.~\ref{sec:calcspectra}.

\subsection{Helicity-dependent splitting of \texorpdfstring{$T$}{T} modes}
\label{sec:splitting}

The triply degenerate $T$ modes behave differently. Figures~\subref{CoSiraman}{c}--\subref{CoSiraman}{f} show enlarged views of individual phonon peaks measured in the two cross-circular configurations. Instead of a large intensity asymmetry, the $T$ modes exhibit a small but systematic shift in peak position between the LR and RL channels, that is, a lifting of the degeneracy between phonon states with opposite CAM. Weak differences in peak intensity are also present for some $T$ modes, beyond the leading-order result $I^{T}_{\mathrm{LR}} = I^{T}_{\mathrm{RL}}$ of Table~\ref{tab:selection}. They are not described by the octupolar channel, which is absent for the $T$ triplet at $\Gamma$ (Sec.~\ref{sec:discussion}). At finite $q$ along $[111]$ the little group is reduced to $C_3$, under which the $T$ triplet decomposes as $A \oplus E^{+} \oplus E^{-}$, the one-dimensional $C_3$ representations with $m = 0$, $+1$, and $-1$. Its $m = \pm 1$ components may then couple to the electronic states with unequal strength, as the $E$ doublet does at $\Gamma$, so that an intensity difference is not forbidden.
 The systematic $T$-mode signature considered below is the shift in peak position.

Table~\ref{tab:splitting} summarizes the measured and calculated peak positions and the resulting splittings $\Delta\omega \equiv \omega_{\mathrm{LR}} - \omega_{\mathrm{RL}}$, with $m$ denoting the crystal-frame CAM of the phonon component selected by each laboratory polarization channel. The sign of the mechanical angular momentum, plotted in Fig.~\subref{dispersion}{b} and listed as AM in Table~\ref{tab:splitting}, follows that of $m$ for $T^{(1)}$ and $T^{(4)}$ but is opposite for $T^{(2)}$. The probed phonon wavenumber is $q \approx 4.0\times10^{5}$~cm$^{-1}$, estimated as $q = 4\pi n/\lambda$ from the real part of the complex refractive index, $n = 2.5$, at 785 nm \cite{Xu2020}. The $T^{(1)}$ and $T^{(2)}$ modes are higher in frequency in the LR configuration, whereas the $T^{(4)}$ mode exhibits the opposite sign. The mode-dependent signs provide an internal validation against a common calibration offset, which would shift all modes in the same direction. The calculated splittings and their comparison with the measurements are discussed in Sec.~\ref{sec:dispersion}.

By contrast, the $E$ modes show no measurable shift in the peak position within the experimental resolution, despite their pronounced intensity asymmetry. As shown in Fig.~\subref{CoSiraman}{e}, the $E^{(2)}$ and $T^{(3)}$ modes overlap, and the two contributions are not separated. Therefore the $T$-mode splittings listed in Table~\ref{tab:splitting} are restricted to isolated peaks. The dominant signatures of the two channels are thus distinguished within a single crystal.

\begin{table*}[t]
\caption{Frequencies, CAM $m$, and helicity-dependent splittings of the triply degenerate $T$ modes of L-CoSi, measured by circularly polarized Raman spectroscopy on the back $\{111\}$ surface and compared with first-principles calculations. The column ``AM'' gives the sign of the angular momentum carried by the phonon eigenmode. Peak positions are rounded to $0.1$~cm$^{-1}$, whereas the splittings $\Delta\omega$ were evaluated from the unrounded fitted positions. The fit uncertainties in $\Delta\omega$ are $0.04$, $0.04$, and $0.21$~cm$^{-1}$ for $T^{(1)}$, $T^{(2)}$, and $T^{(4)}$, respectively. The CAM labels $m$ and the sign convention for $q$ are fixed together by a single choice, taken here so that the LR channel selects $m = +1$. Reversing this choice reverses both, so neither the measured $\Delta\omega = \omega_{\mathrm{LR}} - \omega_{\mathrm{RL}}$ nor its comparison with the calculated values is affected.}
\begin{ruledtabular}
\begin{tabular}{c c c c c c c}
 & \multicolumn{3}{c}{Experiment} & & & Calculation \\
\cline{2-4} \cline{7-7}
Peak & Polarization & Frequency (cm$^{-1}$) & $\Delta\omega$ (cm$^{-1}$) & $m$ & AM & $\Delta\omega$ (cm$^{-1}$) \\
\hline
\multirow{2}{*}{$T^{(1)}$} & LR & 222.6 & \multirow{2}{*}{$+0.3$} & $+1$ & $+$ & \multirow{2}{*}{$+0.2$} \\
 & RL & 222.2 & & $-1$ & $-$ &  \\
\multirow{2}{*}{$T^{(2)}$} & LR & 243.9 & \multirow{2}{*}{$+0.8$} & $+1$ & $-$ & \multirow{2}{*}{$+0.5$} \\
 & RL & 243.1 & & $-1$ & $+$ & \\
\multirow{2}{*}{$T^{(4)}$} & LR & 345.9 & \multirow{2}{*}{$-0.4$} & $+1$ & $+$ & \multirow{2}{*}{$-0.3$} \\
 & RL & 346.3 & & $-1$ & $-$ & \\
\end{tabular}
\end{ruledtabular}
\label{tab:splitting}
\end{table*}

\subsection{Calculated phonon dispersion and CAM}
\label{sec:dispersion}

\begin{figure}
\centering
\includegraphics[width=0.95\hsize]{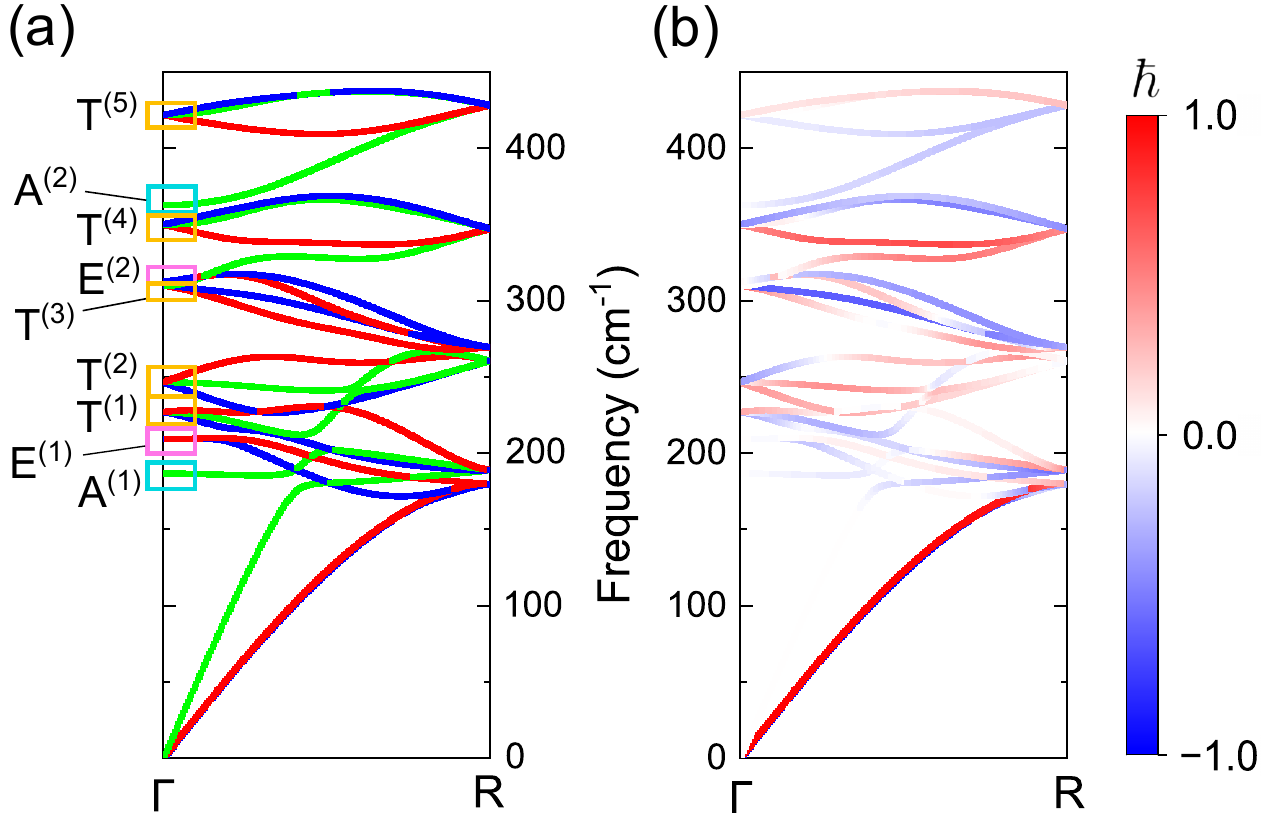}
\caption{(a) Crystal angular momentum $m$ and (b) $[111]$ component of the angular momentum per phonon quantum, in units of $\hbar$, for phonons of L-CoSi along $\Gamma$--$R$. CAM values $m = -1, 0$, and $+1$ are shown in blue, green, and red, respectively.}
\label{dispersion}
\end{figure}

To correlate the observed splittings to the angular-momentum structure of the phonons, we calculated the phonon dispersion of L-CoSi from $\Gamma$ to $R$ (Fig.~\ref{dispersion}), with the branches colored based on their CAM [Fig.~\subref{dispersion}{a}] and their angular momentum along $[111]$ [Fig.~\subref{dispersion}{b}].

Away from $\Gamma$, the triply degenerate $T$ modes split into branches with different CAM, and the branches with opposite CAM disperse in opposite directions, whereas the doubly degenerate $E$ modes show no such splitting at leading order. At $\Gamma$ the $E$ doublet is resolved into two complex components $E^{\pm}$ that create phonons with CAM $m = \pm 1$, but the mechanical angular momentum of each component vanishes identically: the angular momentum is an axial vector, which transforms as $T$ in class $23$, and $E \otimes E$ contains no $T$. A small angular-momentum component is allowed only away from $\Gamma$, where the little group along $[111]$ reduces to $C_3$. Evaluating, for each $T$ mode, the frequency difference between the branches with $m = +1$ and $m = -1$ at the probed wave vector gives the calculated splittings listed in Table~\ref{tab:splitting}. These reproduce the mode-dependent sign change observed experimentally, including the opposite sign of $\Delta\omega$ for $T^{(4)}$ relative to $T^{(1)}$ and $T^{(2)}$, and account for the observed magnitudes to within about $40\%$, the calculated values being systematically smaller. The agreement supports the assignment of the measured circular-polarization-dependent shifts to the CAM-resolved phonon dispersion.

Figures~\subref{dispersion}{a} and \subref{dispersion}{b} also make explicit the distinction between the two quantities \cite{Ishito2023HgS}. For the $T$ branches the lifting of the degeneracy is accompanied by a finite angular momentum, whereas the channel in which a given branch is observed is governed by the CAM $m$, which is defined modulo the discrete rotational symmetry of the lattice. The two need not coincide \cite{Zhang2026}: for $T^{(2)}$ the signs of the CAM and of the angular momentum are opposite (Table~\ref{tab:splitting}), yet it is $m$ that determines which cross-circular channel probes which branch.

\subsection{Calculated circular Raman spectra}
\label{sec:calcspectra}

\begin{figure}
\centering
\includegraphics[width=0.9\hsize]{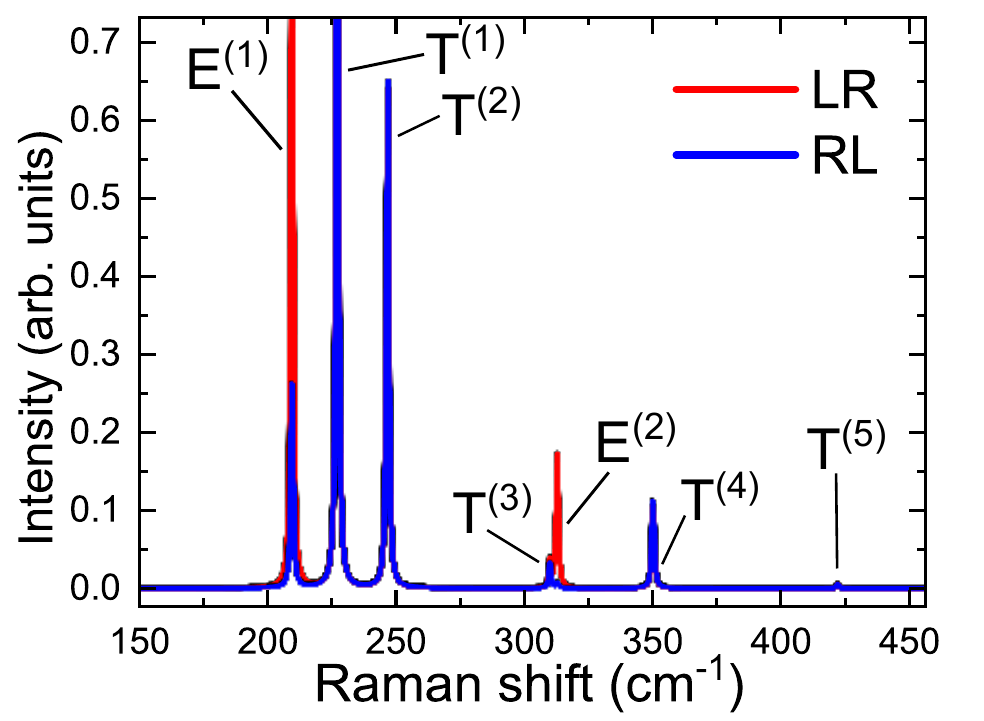}
\caption{First-principles Raman spectra of zone-center phonons of L-CoSi calculated in cross-circular polarization configurations for light propagating along $[111]$ at an excitation energy of 1.20 eV.}
\label{ROA_calculation}
\end{figure}

\begin{figure}
\centering
\includegraphics[width=0.8\hsize]{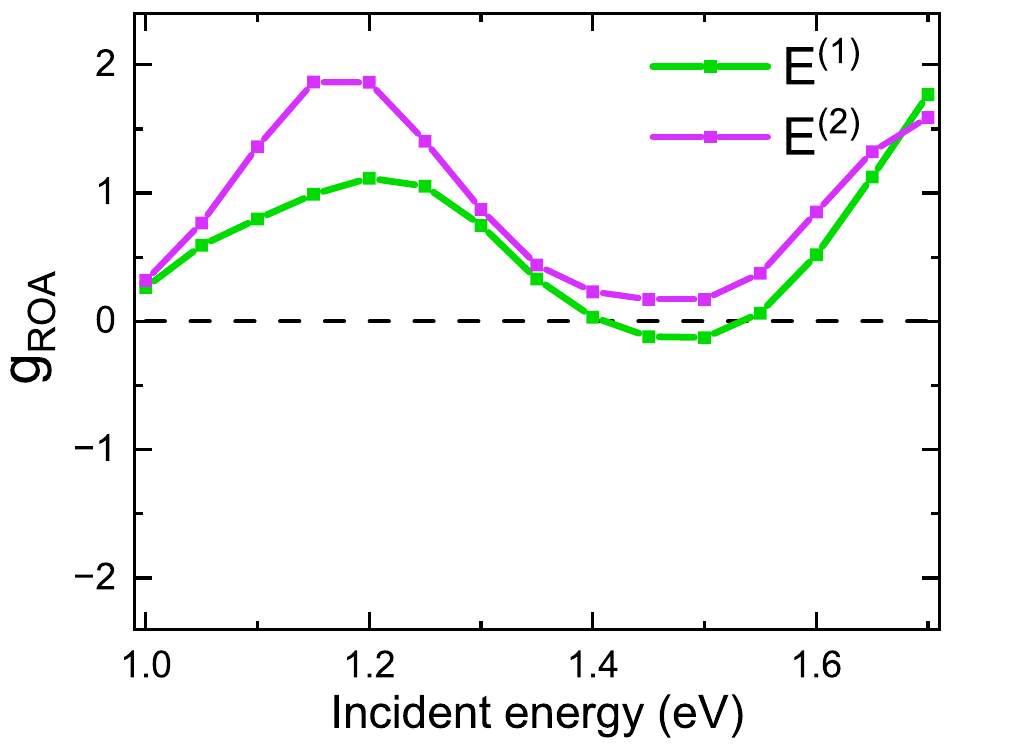}
\caption{Calculated excitation-energy dependence of the ROA factor $g_{\mathrm{ROA}}$ for $E^{(1)}$ and $E^{(2)}$ phonon modes of L-CoSi, for light propagating along $[111]$ and the polarization convention of Sec.~\ref{sec:experiment}.}
\label{fig:ROAenergy}
\end{figure}

Figure~\ref{ROA_calculation} shows the Raman scattering intensities calculated for the cross-circular configurations on the $\{111\}$ plane at an excitation energy of $1.20$ eV, chosen as representative of the resonant regime of the calculations (Fig.~\ref{fig:ROAenergy}). The intensities in Fig.~\ref{ROA_calculation} were calculated for the zone-center phonons, while the finite-$q$ splittings of the $T$ branches were evaluated separately from the phonon dispersion in Fig.~\ref{dispersion}. The calculations yield a clear intensity asymmetry between the two cross-circular polarization channels for the $E$ modes.

For Fig.~\ref{fig:ROAenergy}, $g_{\mathrm{ROA}}$ was evaluated from the calculated mode-resolved $E^{+}$ and $E^{-}$ intensities using Eq.~(\ref{eq:groa}). The calculated ROA depends strongly on the excitation energy, its magnitude showing pronounced enhancement and suppression across the computed range (Fig.~\ref{fig:ROAenergy}). Its sign is positive over that range for the $E^{(2)}$ mode and for the $E^{(1)}$ mode except in a narrow interval near $1.45$--$1.50$~eV. The calculated response is therefore not confined to a single narrow resonance. This supports the resonant character of the response observed at 785 nm ($1.58$ eV) and motivates interpreting the measured wavelength dependence as resonant. 

\section{Discussion}
\label{sec:discussion}

The two circular Raman responses measured above are distinct manifestations of the inequivalence between states with opposite crystal angular momenta, arising through two different symmetry channels, as summarized in Table~\ref{tab:channels}.

In the first channel, which is the one realized by the triply degenerate $T$ modes, the absence of improper rotational symmetry allows $\omega_{+1}(q) \neq \omega_{-1}(q)$ at finite $q$ \cite{Tsunetsugu2026}, where $\omega_m(q)$ is the frequency of the branch with CAM $m$ and $q$ is the signed wave-vector coordinate along $[111]$. This splitting is a property of the phonon dispersion itself rather than of the geometrical relationship between the crystal and the incident light. For the doubly degenerate $E$ modes, by contrast, symmetry forbids a comparable helicity-dependent splitting at leading order in a time-reversal-even crystal \cite{Tsunetsugu2026}, because the antisymmetric $E$-mode bilinear transforms as $A$ whereas the wave vector transforms as $T$, so that no scalar invariant linear in $q$ can be formed. The time-reversal-odd case is discussed below. For the $T$ branches, time reversal imposes $\omega_{m}(q) = \omega_{-m}(-q)$ and thus relates $\omega_{+1}(q)$ to $\omega_{-1}(-q)$, but it does not require $\omega_{+1}(q) = \omega_{-1}(q)$ at the same nonzero $q$. When the opposite $\{111\}$ surface is probed, both $q$ and the crystal-frame CAM selected by a fixed laboratory channel are reversed. The LR and RL channels address the corresponding time-reversed partners at the same frequencies. The helicity-dependent splitting is therefore expected to keep its sign and magnitude on the two opposite surfaces, in contrast to the intensity asymmetry discussed next.

In the second channel, the doubly degenerate $E$ mode is resolved into two complex components $E^{+}$ and $E^{-}$, which create phonons with crystal angular momenta $m = +1$ and $m = -1$, respectively. Their one-phonon normal-coordinate amplitudes and complex Raman susceptibilities are written $\Phi_\pm$ and $\chi_\pm$, with the subscript labeling the phonon created (Appendix), and the two cross-circular channels probe the two components separately.
\begin{equation}
I_{\mathrm{LR}} \propto \left| \chi_+ \Phi_+ \right|^2 ,
\qquad
I_{\mathrm{RL}} \propto \left| \chi_- \Phi_- \right|^2 .
\label{eq:chiphi}
\end{equation}
In the present nonmagnetic crystal, time reversal maps $E^{+}$ onto $E^{-}$ and enforces their degeneracy at $\Gamma$ \cite{Watanabe2025arXiv}. Under the same normal-coordinate normalization, the corresponding one-phonon amplitudes have equal magnitudes: $|\Phi_+| = |\Phi_-| \equiv |\Phi|$. At thermal equilibrium, the occupation factors of the two components are equal. Therefore the observed intensity asymmetry reflects the inequivalence of Raman susceptibilities instead of an occupation imbalance,
\begin{equation}
I_{\mathrm{LR}} - I_{\mathrm{RL}} \propto \left( |\chi_+|^2 - |\chi_-|^2 \right) |\Phi|^2 .
\label{eq:inequivalence}
\end{equation}
It is an axial multipolar degree of freedom that renders $|\chi_+| \neq |\chi_-|$. If time-reversal symmetry were broken by axial multipolar order, the two components would in addition cease to be degenerate: an energy gain linear in the phonon frequency and in $|\Phi_+|^2 - |\Phi_-|^2$ then splits the $E$ doublet \cite{Watanabe2025arXiv, Sutcliffe2026}. This bilinear is distinct from the mechanical angular momentum (Sec.~\ref{sec:dispersion}). The common sign of the Stokes and anti-Stokes intensity asymmetries (Sec.~\ref{sec:roa}) identifies the response as time-reversal even, since the two processes are related by an antiunitary symmetry \cite{Watanabe2025PRB}. The absence of a measurable $E$-mode splitting is consistent with this conclusion. Equivalently, the CAM-resolved amplitudes obtained from the definition in Sec.~\ref{sec:calc} are $M_\pm = \chi_\pm \Phi_\pm$.

\begin{figure}
\centering
\includegraphics[width=0.95\hsize]{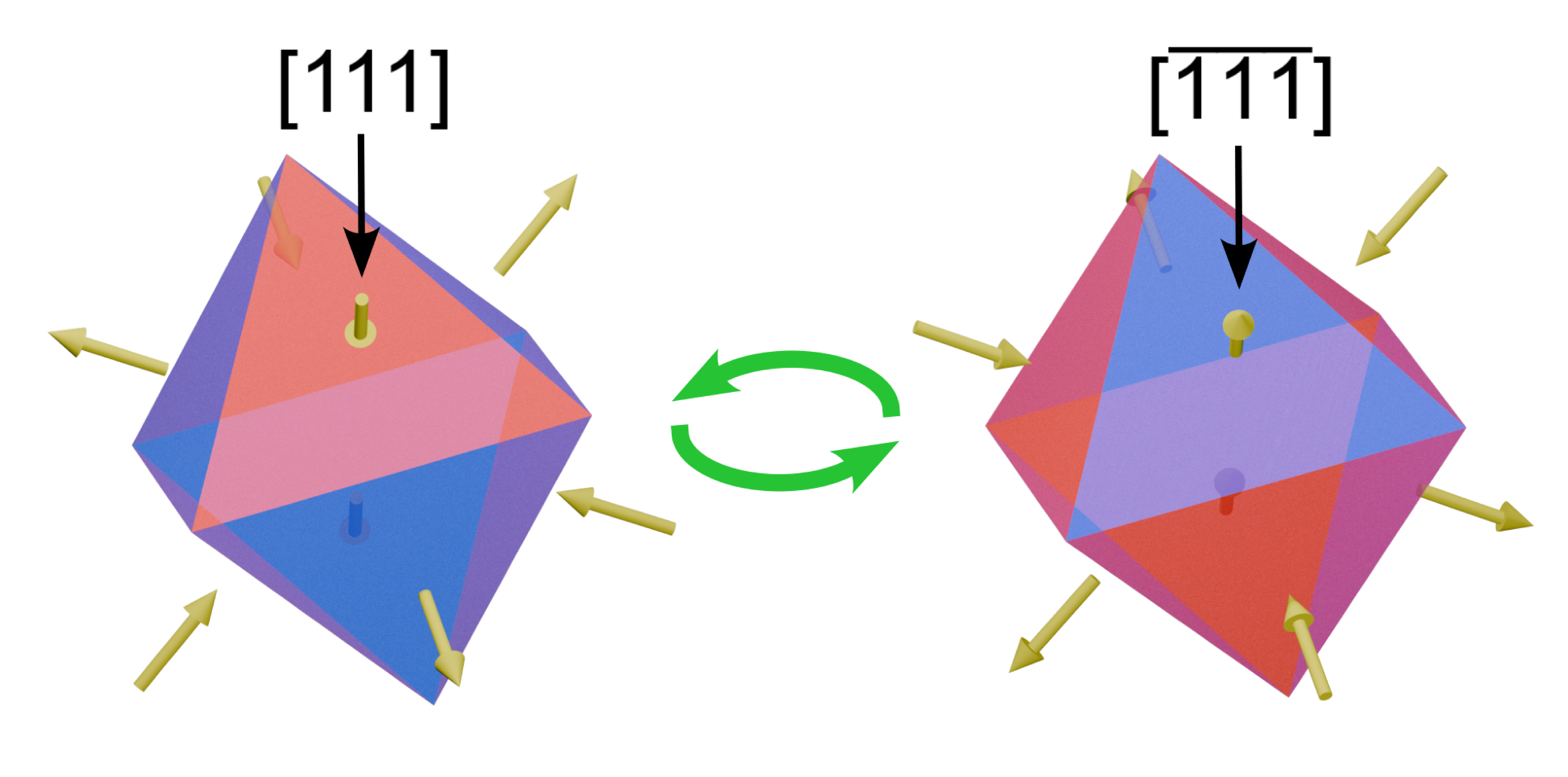}
\caption{Schematic illustration of $xyz$-type electric toroidal octupole viewed along $[111]$ (left) and $[\bar{1}\bar{1}\bar{1}]$ (right). The two views are related by reversal of the viewing direction, $\hat{\mathbf{n}} \to -\hat{\mathbf{n}}$, under which the octupolar form factor $G_{xyz}\, n_x n_y n_z$ is odd. The form factor therefore has opposite signs for the two views, and it vanishes for $\hat{\mathbf{n}} \parallel \langle 110 \rangle$. Yellow arrows represent axial vectors, and octahedron faces are color-coded based on the orientation of these vectors relative to each face. Green arrows indicate that the right panel shows the same octupole viewed from the opposite direction.}
\label{EToct}
\end{figure}

Class $23$ permits an axial electric toroidal octupole $G_{xyz}$ \cite{Hayami2024, Hayami2018} (Fig.~\ref{EToct}). At lowest order in the light propagation direction $\hat{\mathbf{n}} = (n_x, n_y, n_z)$, its contribution to the scattering is proportional to the cubic harmonic $G_{xyz}\, n_x n_y n_z$, which is the lowest-order term that is invariant under the proper rotations of class $23$, odd under $\hat{\mathbf{n}} \to -\hat{\mathbf{n}}$, and identically zero for $\hat{\mathbf{n}} \parallel \langle 110 \rangle$ and $\langle 100 \rangle$. The two opposite faces are not related by any operation of class $23$ (the orbit of $[111]$ under the point group does not contain $[\bar{1}\bar{1}\bar{1}]$). The experimental operation that exchanges the two surfaces may be represented by a proper $180^{\circ}$ rotation of the sample about an axis perpendicular to $[111]$. Polar and axial vectors transform alike under a proper rotation, so the reversal of $\hat{\mathbf{n}}$ relative to the crystal frame, and hence the sign change of the octupolar contribution, does not depend on whether $\hat{\mathbf{n}}$ is taken as the polar propagation direction or as the beam axis associated with the circular-polarization geometry. This realizes, for a chiral cubic class in which every mirror plane is absent, the criterion that the cross-circular response is governed by the breaking of the mirror operation whose plane contains the beam axis \cite{Watanabe2025arXiv}. Two symmetry operations would fix this response to zero: that mirror, and a twofold rotation perpendicular to the beam axis. The latter is a proper rotation that maps $\hat{\mathbf{n}}$ onto $-\hat{\mathbf{n}}$. It therefore forces the part of the cross-circular response that is odd in $\hat{\mathbf{n}}$, which is the parity-even multipolar contribution, to vanish at every order in $\hat{\mathbf{n}}$, while leaving the part that is even in $\hat{\mathbf{n}}$ unconstrained. Class $23$ has no mirror planes, and its twofold axes lie along $\langle 100 \rangle$: none of them is perpendicular to $[111]$, whereas for $\hat{\mathbf{n}} \parallel \langle 110 \rangle$ or $\langle 100 \rangle$ one of them is. The octupolar contribution on $\{110\}$ is therefore forbidden exactly rather than only at lowest order in $\hat{\mathbf{n}}$. It is the spatial parity of the multipole, rather than the mirror symmetry itself, that distinguishes the octupolar from the monopolar contribution discussed below.

In CoSi this role, that of rendering $|\chi_+| \neq |\chi_-|$, is played by the octupole: the leading $G_{xyz}$-mediated antisymmetric-intensity channel is allowed for the $E$ doublet but absent for the $T$ triplet. In class $23$ the self-products of the two mode representations are $E \otimes E = 2A \oplus E$, whose antisymmetric part is the identity representation $A$, to which $G_{xyz}$ belongs, and $T \otimes T = A \oplus E \oplus 2T$, whose antisymmetric part is $T$ and contains no $A$. The cross-circular intensity difference of the $E$ doublet can therefore couple to the $A$-symmetry octupole $G_{xyz}$, whereas at $\Gamma$ the $T$-triplet bilinear cannot couple to $G_{xyz}$ through the same $A$-symmetry channel. Away from $\Gamma$ this restriction is lifted, as discussed in Sec.~\ref{sec:splitting}. The analogous analysis for the $T_h$ point group of pyrite FeS$_2$ is given in Ref.~\cite{Suganuma2026Pyrite}.

Point group $23$ is chiral and therefore also admits a fully symmetric axial invariant of rank zero, the electric toroidal monopole $G_0$ associated with structural chirality \cite{Miki2025}. No axial invariant of rank one or two exists in class $23$: an electric toroidal dipole transforms as an axial vector, which belongs to the $T$ representation, and a rank-two axial invariant would span $E \oplus T$, and neither contains the identity representation, so CoSi is not ferroaxial. The octupole and the monopole are therefore the two lowest-rank candidates, and they differ in spatial parity. The octupole $G_{xyz}$ is parity-even and can contribute already in the wave-vector-independent electric-dipole limit, whereas the monopole $G_0$ is a pseudoscalar and contributes to the odd-parity cross-circular response, whose leading long-wavelength contribution is spatially dispersive and begins at linear order in the wave vector \cite{Watanabe2025PRB}. Reversing the measurement direction exchanges the two cross-circular channels while reversing the relevant wave vectors. Consequently, the front--back difference isolates the even-parity response, whereas the front--back average isolates the odd-parity response. The strong suppression on $\{110\}$, where the leading octupolar form factor $n_x n_y n_z$ vanishes, therefore supports the assignment of the dominant even-parity response to the $xyz$-type octupolar channel.

Similar to the octupolar ROA of pyrite FeS$_2$ \cite{Suganuma2026Pyrite}, this asymmetry does not require a finite mechanical phonon angular momentum. At $\Gamma$, the $E^{\pm}$ components carry the opposite CAM, $m = \pm 1$, whereas their mechanical angular momentum vanishes by symmetry (Sec.~\ref{sec:dispersion}). At the finite wave vector probed here, the calculated $E$ branches carry only $|\mathbf{L}\cdot\hat{\mathbf{n}}| \lesssim 0.01\,\hbar$, which is more than one order of magnitude smaller than that of the $T$ branches [Fig.~\subref{dispersion}{b}]. Therefore the octupole renders the two CAM sectors inequivalent, instead of distinguishing states with opposite mechanical angular momenta.

\begin{table*}[t]
\caption{Two symmetry channels of angular-momentum inequivalence in CoSi, the observable each yields, and the expected behavior of that observable when the same crystal is measured on the opposite $\{111\}$ surface. The sign reversal of the $E$-mode asymmetry follows from the exchange of the two cross-circular channels under reversal of the measurement direction, and its even- and odd-parity components are discussed in Sec.~\ref{sec:discussion}.}
\begin{ruledtabular}
\begin{tabular}{lccc}
Modes & Observable & Origin & Expected front $\leftrightarrow$ back \\ \hline
$E$ & intensity asymmetry & $xyz$-type axial-octupolar channel & reverses \\
$T$ & frequency splitting & finite-$q$ chiral phonon dispersion & unchanged \\
\end{tabular}
\end{ruledtabular}
\label{tab:channels}
\end{table*}

The frequency splitting of the $T$ modes admits a direct interpretation in terms of the topological phonon structure of the B20 compounds, consistent with first-principles studies showing that axial phonons in chiral crystals carry finite angular momentum and that chiral and topological phonon modes can coexist \cite{Reddy2025}. The triply degenerate $T$ modes at $\Gamma$ correspond to the spin-1 nodes predicted for transition-metal monosilicides \cite{Zhang2018}, where the dispersion is described by $H \sim \mathbf{q}\cdot\mathbf{S}$ \cite{Tsunetsugu2026}. The corresponding nodes were observed via inelastic x-ray scattering for the isostructural compound FeSi \cite{Miao2018}. A related connection between Weyl and chiral phonons has been demonstrated in tellurium, where helicity-resolved Raman scattering probes branches with opposite crystal angular momenta \cite{Zhang2023, Ishito2023Te}.

 Raman scattering probes these branches at the small but finite wave vector $q \approx 4.0\times10^{5}$~cm$^{-1}$ set by the real part of the refractive index, where the linear term has lifted the degeneracy, and the cross-circular channels address the branches with $m = +1$ and $m = -1$ separately. Because of the strong absorption at 785 nm ($\kappa = 2.3$ \cite{Xu2020}), the probed wave vector is not sharp but distributed, with a half width at half maximum $\Delta q = 4\pi\kappa/\lambda \approx 3.7\times10^{5}$~cm$^{-1}$ [see Supplemental Material \cite{SM} (Sec.~S1)]. Since the splitting is linear in $q$ near the spin-1 node, $\Delta\omega_i = \alpha_i q$ with a mode-specific slope $\alpha_i$, the Lorentzian wave-vector distribution maps onto a Lorentzian distribution of the splitting, centered at $\alpha_i q_0$ with a half width $|\alpha_i| \Delta q$. Time reversal, $\omega_m(q) = \omega_{-m}(-q)$, makes the mean of the two branches even in $q$, so the individual branch slopes are $v_{\pm,i} = \pm \alpha_i/2$ to linear order in $q$ about $\Gamma$ and each cross-circular peak is broadened by a half width $|\alpha_i| \Delta q/2$. This broadening does not change the mode-dependent signs or the ratios of the splittings evaluated at $q_0$. Deviations from linear dispersion and any $q$ dependence of the Raman matrix elements remain as residual uncertainties, since $\Delta q \sim q_0$.
 Therefore the measured splittings of $0.3$--$0.8$~cm$^{-1}$ provide a helicity-resolved probe of the CAM-resolved branch structure around these nodes, with a peak-position precision substantially finer than that of the inelastic x-ray measurements reported for FeSi \cite{Miao2018}.

The ROA of the $E$ modes has a different character. At leading order, the real, symmetric Raman tensors of class $23$ yield equal susceptibilities for the two CAM components, $\chi_+ = \chi_-$ (Appendix). Therefore the observed asymmetry does not follow from the tensor symmetry alone: it arises through the resonant Raman matrix elements, in which electron--phonon coupling enters and to which the axial multipolar degrees of freedom of the crystal contribute \cite{Watanabe2025arXiv}. The strong dependence on the excitation energy observed in both the experiment and calculations is consistent with this resonant origin. Unlike molecular natural ROA, which requires spatially dispersive contributions such as magnetic-dipole and electric-quadrupole couplings \cite{BarronBook}, the asymmetry in CoSi is allowed within the electric-dipole channel through the resonant matrix elements and axial multipolar symmetry \cite{Watanabe2025arXiv}, which is consistent with the electric-dipole calculations of Sec.~\ref{sec:calc} yielding an $E$-mode intensity asymmetry. The circular Raman response of the $E$ modes is in this sense the more indirect of the two symmetry channels, since it is governed by the scattering matrix elements instead of by the phonon energies.

Taken together, the two symmetry channels place CoSi in a distinctive position with respect to the materials studied so far. Large ROA has been reported for ferroaxial and higher-rank multipolar systems \cite{Kusuno2026NiTiO3, Suganuma2026Pyrite}, helicity-dependent phonon splittings for chiral crystals \cite{Ishito2023HgS, Ishito2023Te, Oishi2024, Kusuno2026CrSi2}, and strong ROA accompanied by calculated chiral phonon modes for layered AgCrP$_2$Se$_6$ \cite{Rao2025}. The two responses have, however, rarely been observed and disentangled in the same material. CoSi hosts both, and they can therefore be compared within a single sample and a single measurement geometry.

\section{Conclusions}

We have shown that chiral CoSi exhibits an intensity asymmetry of the doubly degenerate $E$ modes and a helicity-dependent frequency splitting of the triply degenerate $T$ modes. These two circular Raman responses arise through distinct symmetry channels. The $E$-mode response reflects inequivalent Raman coupling to opposite-CAM states through an axial electric toroidal octupolar channel, whereas the $T$-mode response reflects the finite-$q$ energy splitting of opposite-CAM branches due to structural chirality. Applying the same analysis to enantiomer pairs and across structural or electronic phase transitions could determine whether an angular-momentum-selective response originates from the excitation spectrum or from resonant light-matter coupling, thereby separating the roles of chirality, multipolar order, and the topology of the electronic and phonon bands.

\begin{acknowledgments}
The authors are grateful to H. Kusunose for valuable discussions.
T.S. was supported by JSPS KAKENHI (Grants No. JP21H01032, No. JP22H01154, and No. JP26H02234), MEXT X-NICS (Grant No. JPJ011438), JST CREST (Grant No. JPMJCR24R5).
G.K. was supported by JST SPRING (Grant No. JPMJSP2180) and the Science Tokyo Support Program for Doctoral Students, funded by the Universities for International Research Excellence.
H.W. was supported by JSPS KAKENHI (Grants No. JP23K13058, No. JP24K00581, and No. JP25H02115).
R.O. was supported by JSPS KAKENHI (Grants No. JP26H00618 and No. JP26K17075) and a Special Postdoctoral Researcher Program at RIKEN.
R.A. was supported by JSPS KAKENHI (Grants No. JP25H01246 and No. JP25H01252) and JST CREST (Grant No. JPMJCR23O4).
H.W. and R.A. were supported by RIKEN TRIP initiative (RIKEN Quantum, Advanced General Intelligence for Science Program, Many-Body Electron Systems).
Y.K. was supported by JSPS KAKENHI (Grants No. JP23H01870, No. JP26K01406, and No. JP26H00678).
Y.T. was supported by JSPS KAKENHI (Grants No. JP22H01944, No. JP23H01870, and No. JP23H00091) and JSPS International Joint Research Program (JRP-LEAD with UKRI) (Grant No. JPJSJRP20241710).
T.S. and Y.T. were supported by NINS OML Project (Grant No. OML012301) and JST ERATO (Grant No. JPMJER2503).
\end{acknowledgments}

\appendix*
\renewcommand{\theequation}{A\arabic{equation}}

\section{Raman tensors and polarization selection rules}
\label{app:tensors}

Here, we summarize the Raman tensors of point group $23$ and the selection rules for the $\{111\}$ geometry described in Sec.~\ref{sec:selection}. Except for Eq.~(\ref{eq:appRes}), where the resonant susceptibilities $\chi_\pm(E_{\mathrm{L}})$ are retained, the tensors $R$ used are leading-order real counterparts of the resonant Raman tensor $\hat{\mathcal{R}}_{\nu}(E_{\mathrm{L}})$ introduced in Sec.~\ref{sec:calc}, obtained per unit of the one-phonon normal-coordinate amplitude [Eq.~(\ref{eq:appRfull})].

For point group $23$, the Raman tensors in the cubic coordinate system are \cite{Loudon2001}
\begin{equation}
R(A) = a
\begin{pmatrix}
1 & 0 & 0\\
0 & 1 & 0\\
0 & 0 & 1
\end{pmatrix}
\end{equation}
for the nondegenerate $A$ mode. For the doubly degenerate $E$ mode, the diagonal tensors can be written in cyclic form as follows:
\begin{align}
R(E_x) &= b
\begin{pmatrix}
0 & 0 & 0\\
0 & 1 & 0\\
0 & 0 & 1
\end{pmatrix},
&
R(E_y) &= b
\begin{pmatrix}
1 & 0 & 0\\
0 & 0 & 0\\
0 & 0 & 1
\end{pmatrix},
\nonumber \\
R(E_z) &= b
\begin{pmatrix}
1 & 0 & 0\\
0 & 1 & 0\\
0 & 0 & 0
\end{pmatrix}. &&
\end{align}
Only two are independent because their sum is proportional to $R(A)$. The two-dimensional $E$ representation is obtained from the traceless component. For the triply degenerate $T$ mode,
\begin{align}
R(T_x) &= c
\begin{pmatrix}
0 & 0 & 0\\
0 & 0 & 1\\
0 & 1 & 0
\end{pmatrix},
&
R(T_y) &= c
\begin{pmatrix}
0 & 0 & 1\\
0 & 0 & 0\\
1 & 0 & 0
\end{pmatrix},
\nonumber \\
R(T_z) &= c
\begin{pmatrix}
0 & 1 & 0\\
1 & 0 & 0\\
0 & 0 & 0
\end{pmatrix}. &&
\end{align}
Here, $a$, $b$, and $c$ are real, mode-dependent Raman-tensor coefficients, and the mode indices are suppressed.

Introducing the complex phase $\xi = e^{2\pi\mathrm{i}/3}$ with $1 + \xi + \xi^{-1} = 0$ and setting $C_3$ as the threefold rotation about $[111]$ that cyclically permutes $x \to y \to z \to x$, complex combinations
\begin{align}
R^{(0)}_{\pm} \equiv - \left[ R(E_x) + \xi^{\mp 1} R(E_y) + \xi^{\pm 1} R(E_z) \right]
\end{align}
evaluate to
\begin{align}
R^{(0)}_{\pm} = b\, \mathrm{diag}\!\left(1, \xi^{\mp 1}, \xi^{\pm 1}\right) .
\label{eq:appR0}
\end{align}
The tensor $R^{(0)}_{\pm}$ is associated with the creation of the phonon component $E^{\pm}$, that is, of a phonon with crystal angular momentum $m = \pm 1$. Because the two complex eigenvectors are conjugate partners and the lattice-displacement operator is Hermitian, the normal coordinate whose own $C_3^{+}$ eigenvalue is $\xi$ contains the creation operator of the $m = +1$ phonon, and the one whose eigenvalue is $\xi^{-1}$ that of the $m = -1$ phonon. Accordingly, the index of $\hat{\chi}^{(\pm)}$, $\hat{\mathcal{R}}_\pm$, $\Phi_\pm$, $\chi_\pm$, and $M_\pm$ labels the phonon created in the Stokes process, as does the superscript of $E^{\pm}$, as in Ref.~\cite{Suganuma2026Pyrite}. The LR channel, which creates the $m = +1$ phonon $E^{+}$, therefore carries the amplitude $\chi_{+}\Phi_{+}$.

Equation~(\ref{eq:appR0}) provides the leading-order real-tensor result, in which the Raman response is described by the real, symmetric tensors of class $23$ with a single independent coefficient $b$. In the full resonant response, the common coefficient is replaced by the complex, excitation-energy-dependent Raman susceptibilities $\chi_\pm(E_{\mathrm{L}})$,
\begin{align}
\hat{\chi}^{(\pm)} = \chi_\pm(E_{\mathrm{L}})\, \mathrm{diag}\!\left(1, \xi^{\mp 1}, \xi^{\pm 1}\right) ,
\label{eq:appRes}
\end{align}
with $\chi_+ = \chi_- = b$ at leading order. The symmetry of class $23$ constrains the two eigentensors to this common diagonal form but does not force their complex amplitudes to be equal beyond the leading order. $\Phi_{\pm}$ is the amplitude that multiplies the eigenvector with $m=\mp1$, so that $\Phi_\pm \to \xi^{\pm 1} \Phi_\pm$ under $C_3$. Here, $\hat{\chi}^{(\pm)}$ is the Raman-susceptibility tensor per unit one-phonon normal-coordinate amplitude, $\chi_\pm = (\hat{\chi}^{(\pm)})_{xx}$ are the corresponding scalar susceptibilities, and $\Phi_{\pm}$ are the corresponding amplitudes. Thus the one-phonon Raman tensors introduced in Sec.~\ref{sec:calc} are
\begin{align}
\hat{\mathcal{R}}_{\pm}(E_{\mathrm{L}}) = \Phi_{\pm}\, \hat{\chi}^{(\pm)} .
\label{eq:appRfull}
\end{align}

For circularly polarized light propagating along $[111]$ we use
\begin{align}
\mathbf{e}_\mathrm{L} = \frac{1}{\sqrt{3}}
\begin{pmatrix} 1\\ \xi^{-1}\\ \xi \end{pmatrix},
\qquad
\mathbf{e}_\mathrm{R} = \frac{1}{\sqrt{3}}
\begin{pmatrix} 1\\ \xi\\ \xi^{-1} \end{pmatrix},
\end{align}
and the one-phonon Raman intensity of a component with amplitude $\Phi$ and susceptibility tensor $R$ is $I \propto |\Phi|^2 | \mathbf{e}_\mathrm{s}^{\dagger} R\, \mathbf{e}_\mathrm{i} |^2$. For the two degenerate $E$ components $|\Phi_+| = |\Phi_-| \equiv |\Phi|$, so that $|\Phi|^2$ is common to the two cross-circular channels. Here and in Table~\ref{tab:selection} it is absorbed into the overall intensity scale, as are the corresponding amplitudes of the $A$ and $T$ modes. The two components $E^{+}$ and $E^{-}$ correspond to distinct final phonon states, so their scattering intensities add incoherently, and each cross-circular channel is evaluated with the tensor of the corresponding final state. A direct calculation gives
\begin{align}
\mathbf{e}_\mathrm{R}^{\dagger} \hat{\chi}^{(+)} \mathbf{e}_\mathrm{L} &= \chi_+, &
\mathbf{e}_\mathrm{R}^{\dagger} \hat{\chi}^{(-)} \mathbf{e}_\mathrm{L} &= 0, \nonumber \\
\mathbf{e}_\mathrm{L}^{\dagger} \hat{\chi}^{(+)} \mathbf{e}_\mathrm{R} &= 0, &
\mathbf{e}_\mathrm{L}^{\dagger} \hat{\chi}^{(-)} \mathbf{e}_\mathrm{R} &= \chi_-,
\end{align}
such that the LR and RL configurations couple only to $E^{+}$ and $E^{-}$, respectively. Within the leading real-tensor approximation $\chi_+ = \chi_-$, the two cross-circular configurations probe the two CAM states separately but with equal strength, and the intensity asymmetry arises only from the resonant corrections to $\chi_\pm$ discussed in Sec.~\ref{sec:discussion}. Thus the quantities $\chi_\pm$ are the polarization projections of the susceptibility tensors, whereas the full CAM-resolved one-phonon scattering amplitudes are $M_\pm = \chi_\pm \Phi_\pm$, which aligns with Sec.~\ref{sec:discussion} and the Supplemental Material \cite{SM} (Sec.~S2). The components $E^{\pm}$ used here are the $E_g^{\pm}$ of Ref.~\cite{Suganuma2026Pyrite}, and $\hat{\chi}^{(\pm)}$, $\Phi_\pm$, $\chi_\pm$, and $M_\pm$ follow the same convention as in that work.

For the $A$ mode one obtains $I^{A}_{\mathrm{LR}} = I^{A}_{\mathrm{RL}} = 0$ and $I^{A}_{\mathrm{LL}} = I^{A}_{\mathrm{RR}} = a^2$. For the $T$ modes the three degenerate components correspond to distinct final states, so their scattering probabilities add incoherently, $I^{T} \propto \sum_{\mu} | \mathbf{e}_\mathrm{s}^{\dagger} R(T_\mu) \mathbf{e}_\mathrm{i} |^2$. Using $\mathbf{e}_\mathrm{R}^{\dagger}R(T_x)\mathbf{e}_\mathrm{L} = 2c/3$, $\mathbf{e}_\mathrm{R}^{\dagger}R(T_y)\mathbf{e}_\mathrm{L} = 2c\xi/3$, and $\mathbf{e}_\mathrm{R}^{\dagger}R(T_z)\mathbf{e}_\mathrm{L} = 2c\xi^{-1}/3$, one finds
\begin{align}
I^{T}_{\mathrm{LR}} = I^{T}_{\mathrm{RL}} = \frac{4}{3}c^2,
\qquad
I^{T}_{\mathrm{LL}} = I^{T}_{\mathrm{RR}} = \frac{1}{3}c^2 ,
\end{align}
such that the parallel-circular intensities are one-quarter of the cross-circular intensities. Decomposing the $T$ modes into CAM eigenstates $m = 0, \pm 1$ yields the same total intensity. The results are presented in Table~\ref{tab:selection}.

For light propagating along $\langle 110 \rangle$, the beam axis is not a threefold axis of point group $23$. Consequently, the CAM label defined with respect to $[111]$ does not provide the circular-polarization selection rule in Table~\ref{tab:selection}. In particular, the two cross-circular channels no longer separately select the $m = \pm 1$ components.

\bibliography{refs}

\end{document}

%% file: revpackage.tex
\usepackage{graphicx}

\usepackage{bm}
\usepackage{xcolor}
    
\usepackage{amsmath, amssymb, amsfonts}

\usepackage[colorlinks=true,allcolors=blue]{hyperref}

\usepackage{siunitx} 
\NewCommandCopy\sqty\qty 
\NewDocumentCommand\Sqty{smm}{%
    \IfBooleanTF{#1}%
    {\sqty[quantity-product={}]{#2}{#3}}%
    {\sqty[quantity-product=~]{#2}{#3}}%
}
\usepackage[italicdiff]{physics} 
\usepackage[version=4]{mhchem} 

\usepackage{cleveref}
\crefname{equation}{Eq.}{Eqs.} 

\NewDocumentCommand\qtxt{sm}{%
    \IfBooleanTF{#1}%
    {~\textrm{#2}}%
    {\quad\textrm{#2}}%
}

\NewDocumentCommand{\subref}{mm}{\hyperref[#1]{\ref{#1}(#2)}}
\graphicspath{%
{../fig/pdf/}%
}

\usepackage{braket}